# Deep-Learning-based Frequency-Domain Watermarking for Energy System Time Series Data Asset Protection

Zhenghao Zhou [a,b,c], Yiyan Li [a,b,c*], Xinjie Yu [a,b,c], Jian Ping [a,b,c], Xiaoyuan Xu [b,c], Zheng Yan [b,c], Mohammad Shahidehpour [d]

[a] *College of Smart Energy, Shanghai Jiao Tong University, Shanghai, 200240, China*
[b] *The Key Laboratory of Control of Power Transmission and Conversion, Ministry of Education, Shanghai Jiao Tong University, Shanghai 200240, China*
[c] *Shanghai Non-Carbon Energy Conversion and Utilization Institute, Shanghai Jiao Tong University, Shanghai 200240, China*
[d] *Galvin Center for Electricity Innovation, Illinois Institute of Technology, Chicago, IL 60616 USA*

**Abstract:** Data has been regarded as a valuable asset with the fast development of artificial intelligence technologies. In this paper, we introduce deep-learning neural network-based frequency-domain watermarking for protecting energy system time series data assets and secure data authenticity when being shared or traded across communities. First, the concept and desired watermarking characteristics are introduced. Second, a deep-learning neural network-based watermarking model with specially designed loss functions and network structure is proposed to embed watermarks into the original dataset. Third, a frequency-domain data preprocessing method is proposed to eliminate the frequency bias of neural networks when learning time series datasets to enhance the model performances. Last, a comprehensive watermarking performance evaluation framework is designed for measuring its invisibility, restorability, robustness, secrecy, false-positive detection, generalization, and capacity. Case studies based on practical load and photovoltaic time series datasets demonstrate the effectiveness of the proposed method.

## 1. Introduction

In recent years, with the fast development of artificial intelligence (AI) technologies, datasets are progressively regarded as valuable assets for supporting data-driven research and applications [1]. In such circumstances, data has been increasingly shared or trated across different entities to maximize its value. For example, the United Nations has called for establishing a global data governing framework to facilitate cross-border data flows, whereas China has proposed the construction of a comprehensive data system aimed at removing institutional barriers during data sharing and trading. However, during the data sharing or trading process, issues such as data security, privacy and copyright protection start to emerge, especially for energy datasets which are highly correlated with energy security and user privacy [2], [3].

Among these challenges, copyright protection (or called *ownership verification*) is gaining increasing attention. As a kind of virtual digital asset, when the energy datasets are being shared or traded, they suffer from the risk of being illegally distributed or sold to third parties by unauthorized individuals for personal gains without informing the dataset owner. Such actions are covert and hard to prevent, causing huge potential losses for the dataset owner. As such, it is important for the official data exchange platforms, such as data trading center, to introduce both new technical tools and management mechanisms to protect the legitimate rights and interests of data owners against infringement, misuse, or tampering.

Several existing techniques can be used for dataset protection, such as encryption [4], hash [5] and differential privacy [6]. Note that these techniques have different purposes and functionalities. Encryption relies on cryptographic algorithms and secret keys to transform plaintext into unintelligible ciphertext, which aims to ensure data confidentiality and prevent unauthorized access to the dataset. For example, Sun et al. employ asymmetric homomorphic encryption to achieve coordinated optimization while preventing unauthorized access [7]. Hash-based methods employ a one-way cryptographic function to map arbitrary-length data to a fixed-length hash value, focusing on verifying data integrity rather than establishing data ownership. Slight modifications to the dataset would lead to drastic changes to the hash encoding, making such modifications noticeable. Tian et al. leverage the Merkle tree structure in conjunction with hash values to verify data integrity and detect as well as locate false data injection attacks by comparing root hash values [8]. Differential privacy introduces controllable noises to the datasets to make the user-specific information unidentifiable, so that the user privacy behind the datasets can be protected [9]. However, all the above techniques cannot help verifying dataset ownership. Once the datasets are released (e.g., distributed by malicious users or password leakage), one can make arbitrary use of the datasets without informing or acknowledging the original dataset owner. In other words, from the perspective of the dataset owner, the question of "*how to prove a public dataset belongs to me*" has not been well answered.

Aiming at this question, *watermarking* has been proven as an effective way of maintaining copyrights of open-access digital assets, such as images, audios, texts, etc. By embedding either visible or invisible information into original files, watermarking technology can ensure authenticity of the digital assets and help the asset owner verify the ownership when necessary [10], [11]. For instance, model-based methods are applied to image watermarking in the spatial domain [12], frequency domain [13] and wavelet domain [14]. These methods embed watermarks by replacing minor components of the original data, introducing explicit numerical modifications and exhibiting limited robustness, as the embedded watermark becomes difficult to detect upon subsequent data alterations. With the development of deep learning techniques, neural networks can learn to embed tiny perturbations into digital assets as invisible watermarks based on encoder-decoder frameworks [15]. For image watermarking, encoders take both the image and the watermark as inputs to generate a watermarked image, while decoders attempt to extract the watermark from the watermarked image [16]. Compared to model-based watermarking techniques, deep learning methods demonstrate higher invisibility, stronger robustness, larger encoding capacity, and excellent real-time decoding capabilities [17], [18], [19].

In energy systems, the advantages of watermarking technology have gained increasing attention. Table I summarizes existing studies of watermarking applications in the energy system domain. It can be noticed that existing studies mainly focus on scenarios such as cyber-attack detection and data authentication. For instance, [20] proposes a dynamic watermarking method that can detect cyberattacks by embedding random signals into photovoltaic control systems, while [27] introduces a timestamp-based digital text watermarking technique to detect data integrity and replay attacks on energy systems.

However, from the perspective of data asset protection, the study of watermarking technology is still at the infant stage with few literatures found. The methods in existing studies cannot be directly implemented to achieve data asset ownership verification due to the different requirements for watermarking characteristics: Existing applications primarily aim to ensure data integrity and real-time security, for which watermarks are designed to be fragile such that any minor modification will make the watermark undetectable and triggers the alert. In contrast, data asset ownership protection requires a highly robost watermark, where watermarks must remain detectable and extractable even after the data has undergone malicious tampering or noise attacks.

Moreover, the unique characteristics of energy data impose stringent requirements on the "invisibility" of the embedded watermarks. This is because the energy system data has clear and and critical physical meanings, where even minor numerical perturbations may correspond to significant changes in system operating conditions.Therefore, the corresponding methodology and evaluation criteria for implementing watermarking to assert energy data asset ownership still need to be explored.

TABLE I Summary the Watermark Applications in Energy System

| **Scenario** | **Description** | **Entity** | **Watermark detection** | | **Objective** |
|---|---|---|---|---|---|
| | | | **Original data** | **Malicious modification** | |
| **Attack detection and data authentication** | Embeds a specific watermark to ensure data integrity and detect malicious attacks. | Control Signals[20], [21], [22], [23], [24], [25] | Yes | No | Ensure cyber security and verify data integrity |
| | | System data [26], [27], [28] | Yes | No | |
| **Ownership verification (the proposed method)** | Embeds robust and invisible watermarks to provide verifiable copyright claims for data owners. | Time Series Datasets | Yes | Yes | **Verify ownership to protect data asset copyright** |

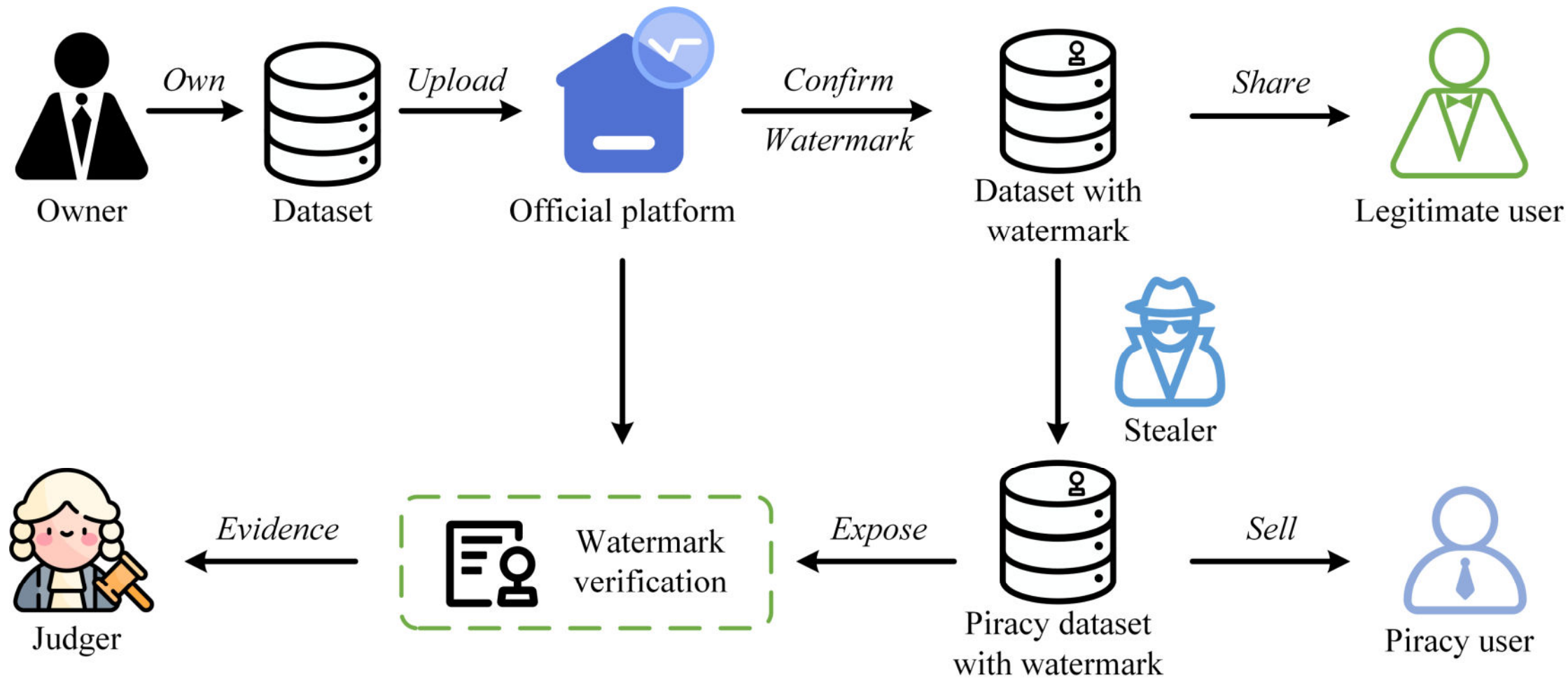

Fig. 1. The scenario of dataset ownership verification. As a certificate, watermark can prove ownership.

In this paper, we propose a frequency-domain deep-learning watermarking method for asserting energy system time-series data asset ownership. By designing specialized network structure and loss function, the proposed method can embed watermarks into the original dataset. Fig. 1 demonstrates an application scenario of the proposed watermarking method. When an owner uploads his dataset to an official data sharing/trading platform, a unique watermark will be embedded into the dataset for ownership verification. In the event of any subsequent copyright disputes, the official authority can determine data ownership by recognizing this watermark. For example, once the illegal piracy dataset is exposed, the owner can request the platform to assert his ownership to this dataset by extracting the predefined watermark within the dataset and take legal actions to the stealer/piracy user when necessary.

The main contributions of our paper are considered twofold:

1) The watermarking concept is introduced to the energy sector to achieve ownership verification of the energy system time series data assets. With only negligible modifications to the original data, the watermarks are invisible to make them only detectable by the watermark maker, and are robust enough to endure severe data distortions.

2) A deep-learning based frequency-domain watermarking model is proposed. The model is formulated as an encoder-decoder framework with specially designed loss functions. Particularly, a frequency-domain data preprocessing method is designed to eliminate frequency bias when training neural networks on time series datasets.

The rest of the paper is organized as follows: Section II introduces the basic concepts, Section III presents the watermarking methodology, Section IV demonstrates the case study results, Section V discusses the performance and limitation, and Section VI concludes the paper.

## 2. Basic concepts

In this section, we introduce watermarking and frequency bias before discussing the proposed watermarking method.

### 2.1 Watermarking

Watermarking is a form of steganography that embeds additional information into digital assets in either visible or invisible way and can be detected and extracted by the owner when necessary. Watermarks can be categorized into three types based on extractability [29]:

**Blind:** The extraction of blind watermarks requires only the watermarked data without relying on original data or information on watermarking process.

**Semi-blind:** The extraction of semi-blind watermarks requires watermarked data plus information on watermarking process.

**Non-blind:** The extraction of non-blind watermarks requires complete information on watermarked data, original data, and watermarking process.

Blind watermarking is the most challenging but also the most practical solution because users can only access the watermarked data in practice. As a result, blind watermark is designed in this paper.

For data asset protection, a successful watermark must have the following key characteristics [29], [30], [31]:

**Invisibility:** Watermarking should neither alter the characteristics nor influence the quality of original data. Otherwise, watermarking is considered a noise added to the original dataset and deemed unsuccessful.

**Restorability:** The watermarking system must ensure that the embedded watermark can be accurately restored by the owner. The extracted watermark should be highly analogous to the embedded one with minimal distortion.

**Robustness:** Watermarking should be resilient to any post-processing or malicious tampering, such as removal or alteration. Even as the data undergoes compression, transformation or noise injection, watermarking must remain detectable and intact. This ensures that watermarking can endure practical scenarios without degradation.

**Low false-positive detection:** The probability of false positive detection should be exceedingly low in watermarking, indicating that the data without watermarking would not be identified as containing one.

**Generalization:** An efficient watermarking framework should be adaptive to different types of datasets without any performance degradation.

**Secrecy:** Watermarking must be sufficiently secure so that unauthorized parties cannot detect or extract the process. Third parties identifying a watermark, may attempt to manipulate or remove it.

**Capacity:** Watermarking should bear a sufficiently large information capacity which can embed an effective protection process.

## 2.2 Frequency Bias of Neural Networks

Time series can be decomposed into sub-series with varying frequencies and amplitudes. Recent studies indicate that when directly learning from time series, neural networks tend to resort to "*frequency bias*" which focus on components with lower frequency and higher magnitude while being insensitive to high-frequency components [32], [33], [34], [35]. Such characteristics could influence the model accuracy when solving delicate tasks like watermarking. To address this issue, we use two small cases to demonstrate the frequency bias phenomena, which guide the frequency-domain deep-learning model design in the next section.

Denote the original time series data as $\mathbf{X} = \{x_1, x_2, \ldots, x_L\}$, where $L$ is the length of time series. As a discrete variable, we analyze the frequency components of $\mathbf{X}$ using Discrete Fourier Transform (DFT). Then the frequency domain data can be reconstructed in time domain by Inverse Discrete Fourier Transform (IDFT):

$$a_k = \frac{1}{L}\sum_{n=0}^{L-1} x_n e^{-\frac{2i\pi kn}{L}} \tag{1}$$

$$x_n = \sum_{k=0}^{L-1} a_k \cdot e^{\frac{2i\pi kn}{L}} \tag{2}$$

where $x_n$ is the $n$-th sample of the time series, $a_k$ represents the transformed result for frequency $k$, indicating the complex value of the $k$-th frequency component in the sequence. The DFT coefficients $\mathbf{A} = \{a_1, a_2, \ldots, a_L\}$ represent the amplitude.

We employ Fourier analysis using the relative error $\Delta_k$ stated in (3) to measure the convergence behavior across components with different frequencies and amplitudes during training, aiming to quantify the frequency bias.

$$\Delta_k = \frac{\left|a_k^h - a_k^f\right|}{\left|a_k^f\right|} \tag{3}$$

where $k$ represents the frequency, $a_k^h$ and $a_k^f$ respectively are the DFT results of the neural network output and the ground truth, and $|\cdot|$ denotes the norm of a complex number [24][27]. Next, we design two small cases to help better understand the frequency bias concept.

### 1) *Bias to Frequency*

We design a small case where a two-layer neural network fits three sine functions with different frequencies. Fig. 2 shows the time- and frequency-domain fitting results after 1000 training epochs. Table II presents the corresponding fitting error. The neural network performance varies significantly even under the same training set up, showing evident preferences for low-frequency components.

### 2) *Bias to Amplitude*

Based on the same two-layer neural network, another toy case is designed to demonstrate the neural network bias regarding amplitude. In this case, three signals are generated by combining sine functions with varying amplitudes, shown as Fig. 3. The fitting accuracy after 1000 training epochs are shown in Table III, where the neural network tends to well fit the frequency component with larger amplitude.

Table II Relative Error of Different Frequencies Functions

| Frequency | 1Hz | 5Hz | 10Hz |
|---|---|---|---|
| Relative error | 0.0005 | 0.4431 | 0.8395 |

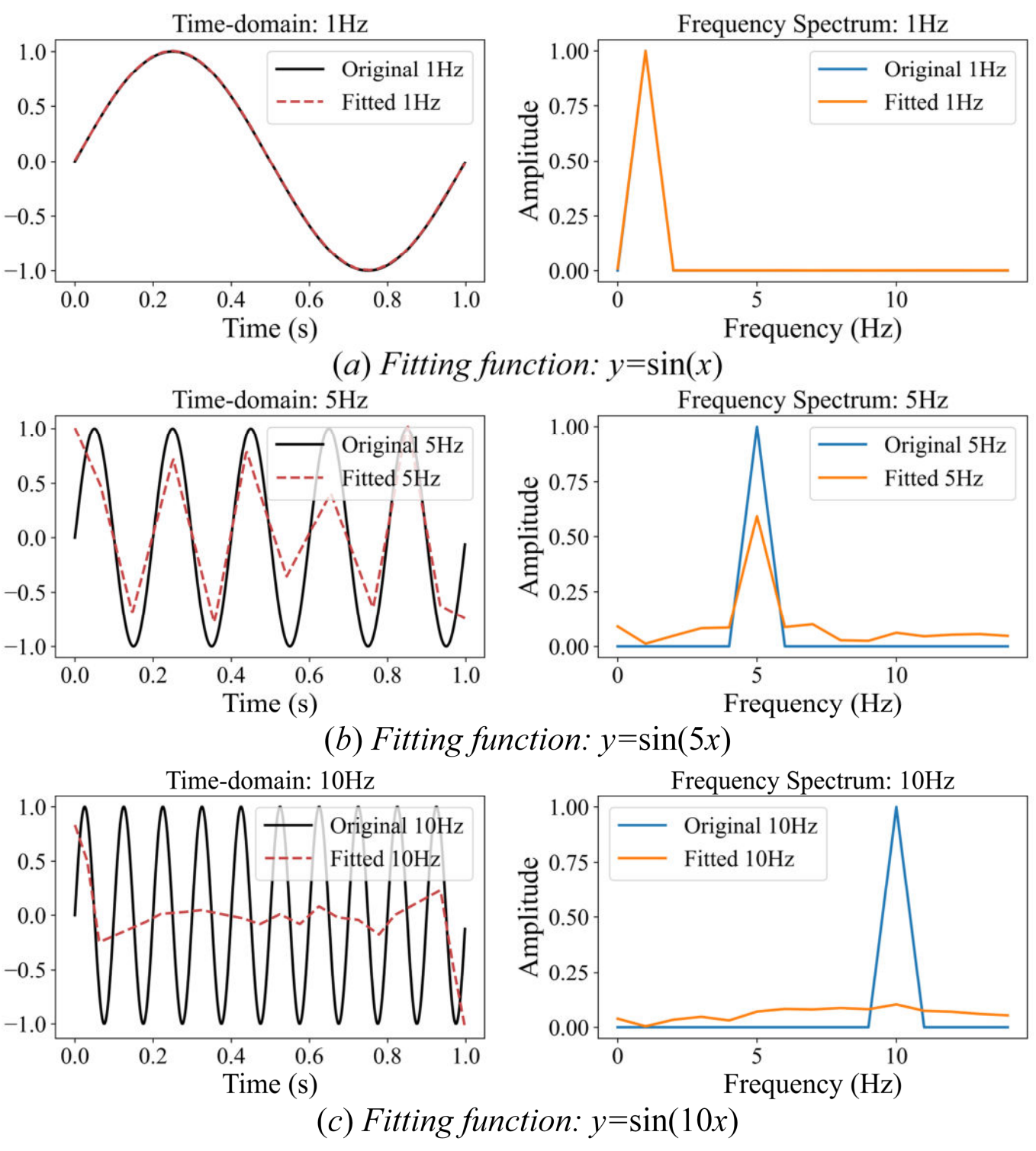

(*a*) *Fitting function:* $y=\sin(x)$

(*b*) *Fitting function:* $y=\sin(5x)$

(*c*) *Fitting function:* $y=\sin(10x)$

Fig. 2. Comparison of neural network fitting results for sine signals with different frequencies. The left column shows the time-domain comparisons, and the right shows the frequency-domain comparisons.

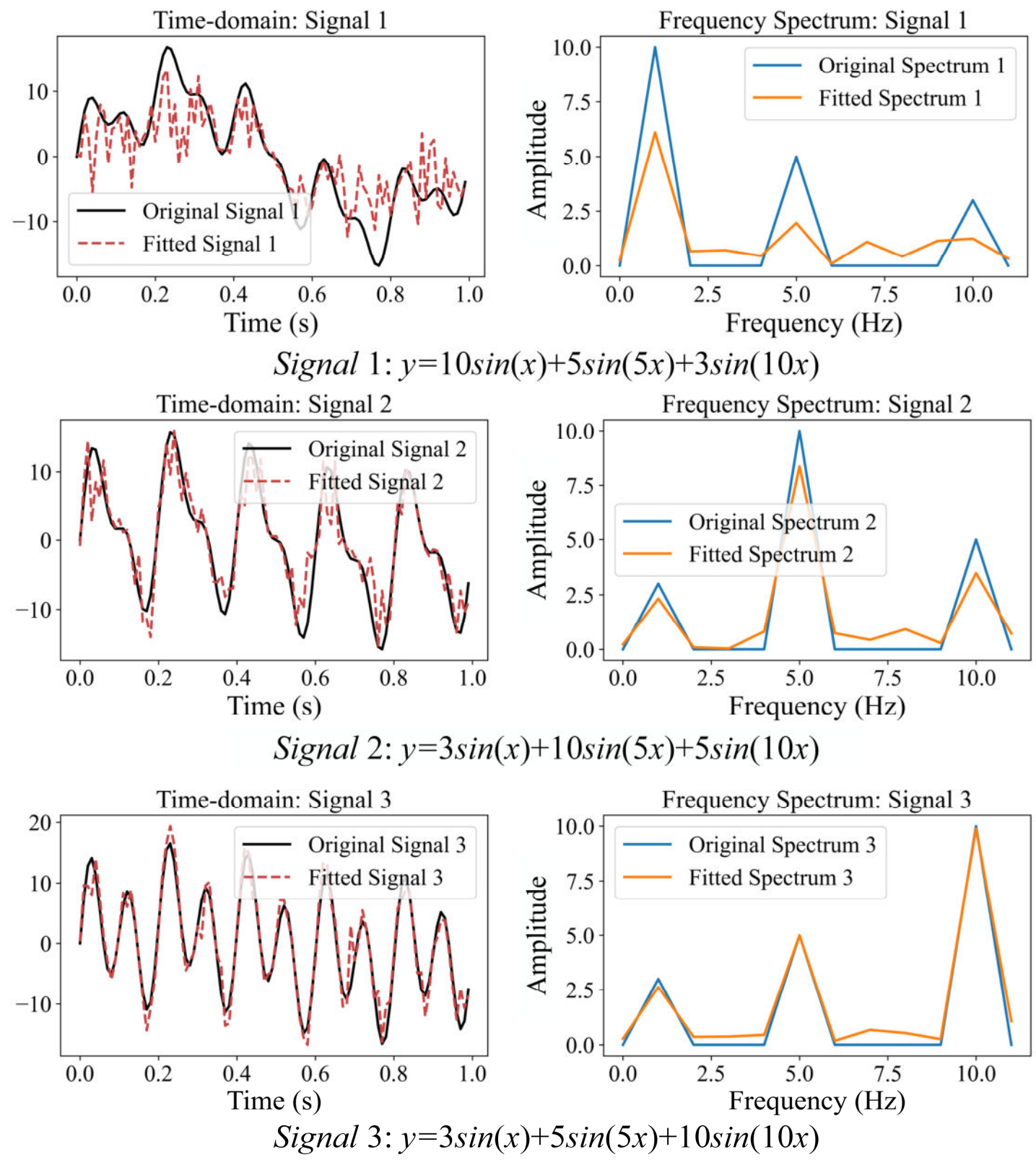

*Signal* 1: $y=10sin(x)+5sin(5x)+3sin(10x)$

*Signal* 2: $y=3sin(x)+10sin(5x)+5sin(10x)$

*Signal* 3: $y=3sin(x)+5sin(5x)+10sin(10x)$

Fig. 3. Comparison of neural network fitting results on signals with varying frequency-domain amplitudes. The left shows the time-domain comparisons, and the right shows the frequency-domain comparisons.

Table III Relative Error of Different Amplitudes Signals

| Primary frequency | Relative error | | |
|---|---|---|---|
| | $\Delta_1$ (1Hz) | $\Delta_5$ (5Hz) | $\Delta_{10}$ (10Hz) |
| $a_1 > a_5 > a_{10}$ | 0.0236 | 0.0657 | 0.2074 |
| $a_5 > a_{10} > a_1$ | 0.1958 | 0.1746 | 0.3808 |
| $a_{10} > a_5 > a_1$ | 0.4441 | 0.4142 | 0.2821 |

## 3. Methodology

In this section, we propose a frequency-domain, deep-learning based watermarking framework that satisfies the watermark characteristics and is free from the frequency bias issue. The proposed watermarking model is shown in Fig. 4, following an encoder-decoder architecture. Denote the encoder and decoder networks as E and D, respectively. The model inputs include both the frequency-domain data $\mathbf{x}_o$ after preprocessing and the watermark $w$ to be embedded. The watermark can be text strings with clear meanings, which will be converted to bitstrings $w \in \{0,1\}$m by ASCII encoding (see Section 3.4 for details).

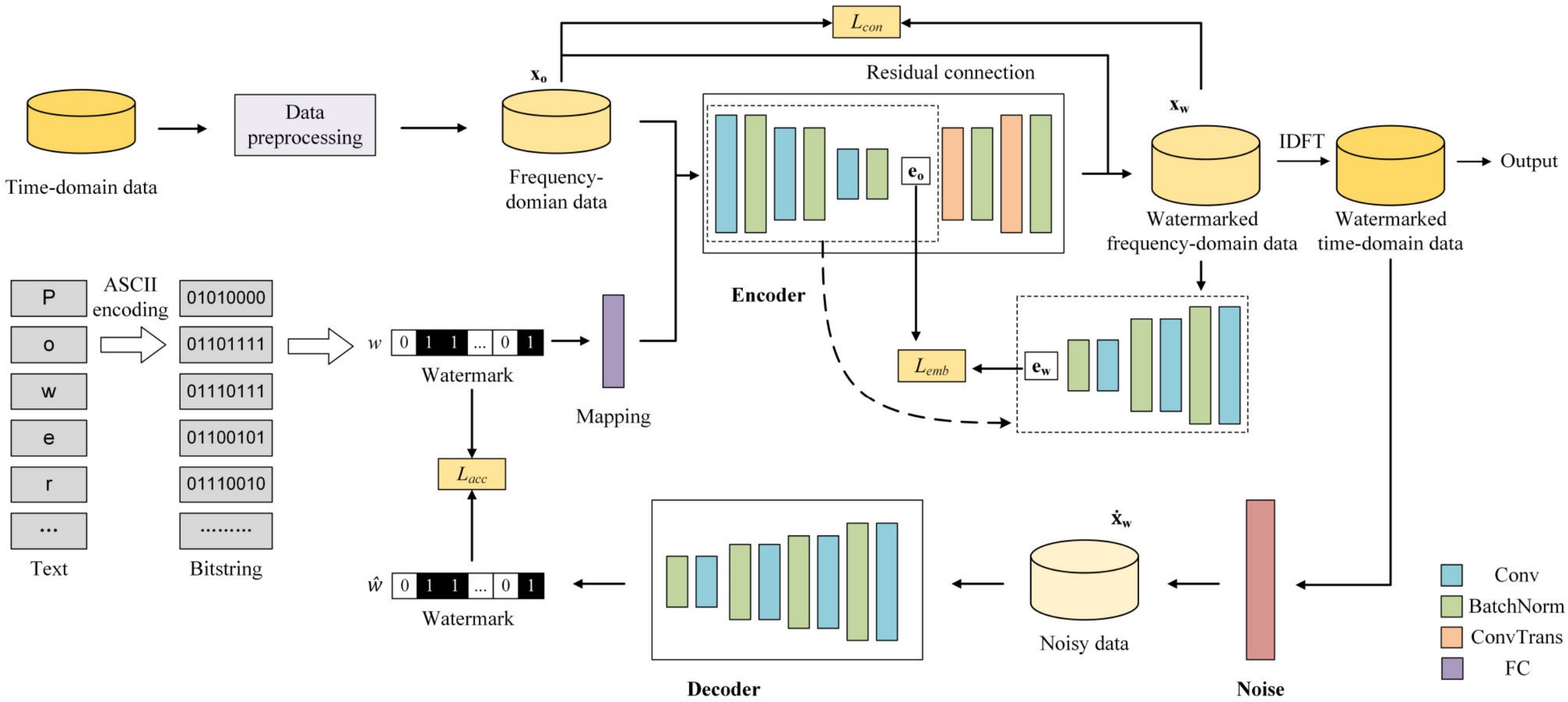

Fig. 4. The proposed watermarking framework

The input $\mathbf{x}_o$ and $w$ are first processed together through convolutional layers to reduce dimensionality, resulting in an intermediate embedding $\mathbf{e}_0$. Then the watermarked data $\mathbf{x}_w$ is produced through transposed convolution. To retain more original feature information, a residual connection is introduced between the input and output of encoder. At this stage, the watermarked data remains in the frequency domain and will be transformed back to the time domain via IDFT. Note that a noise layer is introduced between the encoder and the decoder to simulate the real-world data disturbances to enhance the watermarking robustness (see Section 3.3 for details). After noise is added to $\mathbf{x}_w$, noted as $\dot{\mathbf{x}}_w$, the decoder tries to recover $w$ from $\dot{\mathbf{x}}_w$, which is noted as $\hat{w}$.

### 3.1 Test Case Setup Frequency-Domain Data Preprocessing

The frequency-domain data preprocessing workflow is summarized in Fig. 5. For the original time series dataset $\mathbf{X} \in \mathbb{R}^{N \times L}$, where $L$ is the length of the time series and $N$ is the number of features, we apply DFT to convert the data from the time domain to the frequency domain. By taking only the positive frequency components, we obtain the frequency-domain matrix $\mathbf{A} \in \mathbb{R}^{N \times K}$, and $K$=0.5$N$+1.

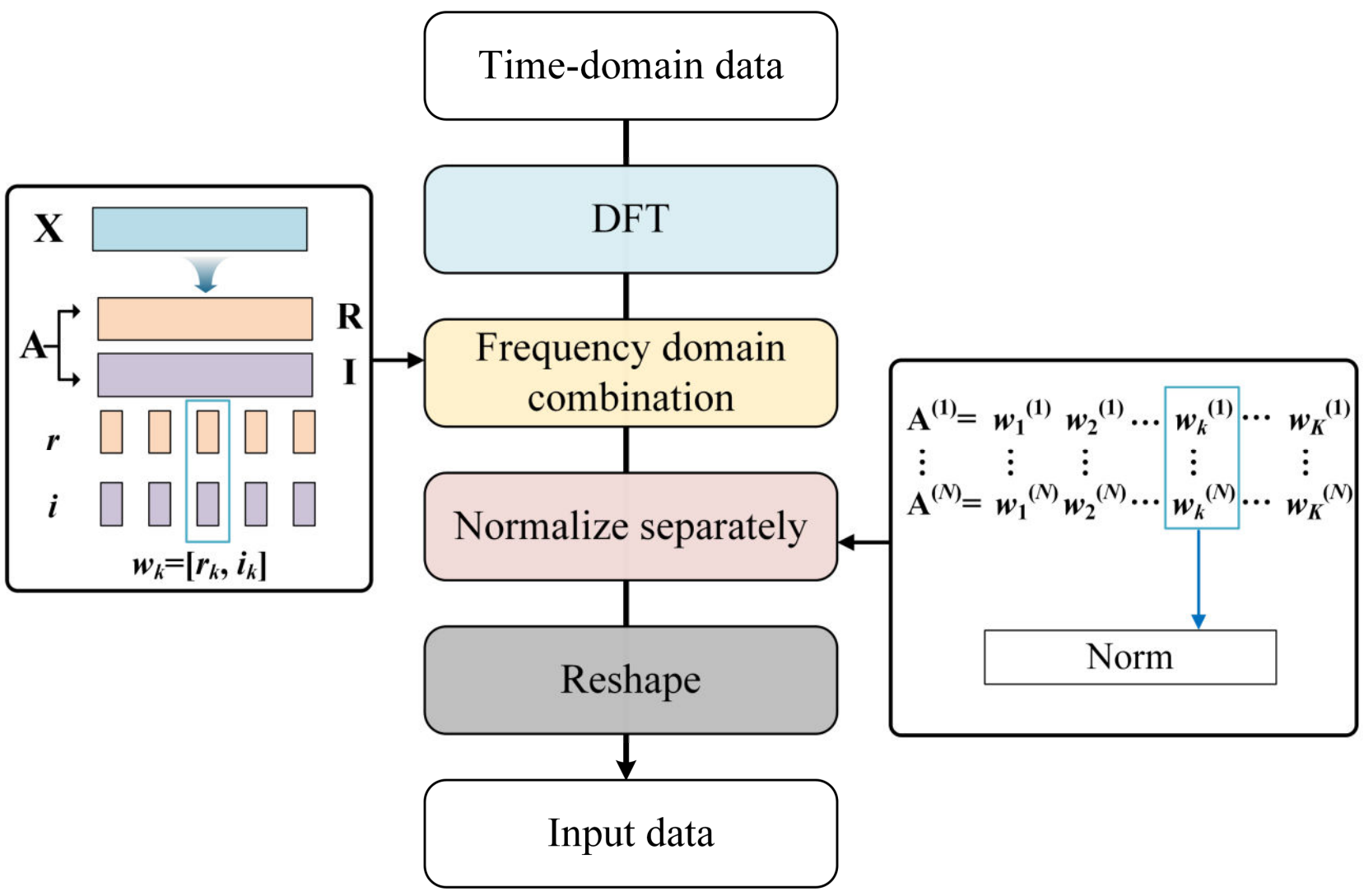

Fig. 5. The data preprocessing workflow.

$$\mathbf{A} = \mathrm{DFT}(\mathbf{X}) = \begin{bmatrix} a_1^1 & a_2^1 & \cdots & a_K^1 \\ a_1^2 & a_2^2 & \cdots & a_K^2 \\ \vdots & \vdots & \ddots & \vdots \\ a_1^N & a_2^N & \cdots & a_K^N \end{bmatrix} \tag{4}$$

$$a_k^n = r_n^k + j i_n^k \tag{5}$$

In practice, **A** is a complex matrix that requires processing by specialized complex-valued neural networks. To simplify the computation, we separate the real and imaginary parts, obtaining a real part matrix $\mathbf{R} \in \mathbb{R}^{N \times K}$ and an imaginary part matrix $\mathbf{I} \in \mathbb{R}^{N \times K}$.

We normalize the frequency-domain values for all samples at the same frequency and then reshape the normalized samples from a one-dimensional sequence into a two-dimensional matrix, before being fed into the watermarking model. After embedding the watermark into the frequency components, we apply IDFT to transform the data back to the time domain. Such data preprocessing approach effectively mitigates the frequency bias in the follow-up neural network training process.

- **Normalization of Frequency-Domain Values**: By normalizing the frequency-domain values for all samples, the bias to amplitude can be avoided.
- **Reshaping into a Two-Dimensional Matrix**: By reshaping the frequency-domain data into a two-dimensional matrix, components with different frequencies can be learned simultaneously by the neural network so that the bias to frequency can be avoided.

### 3.2 Loss Function Design

The training process of encoder and decoder is synchronized with specially designed loss functions. To ensure the invisibility of watermark, the encoder's objective is to generate the watermarked data $\mathbf{x_w}$ that is as close as possible to the original data $\mathbf{x_o}$. On the other hand, the decoder's goal is to accurately reconstruct the watermark $w$, enhancing the model's overall accuracy. Therefore, the loss functions during the training phase are defined as follows:

$$\begin{aligned} \min_{E,D} \mathbb{E}_{\mathbf{x_o} \sim \mathbb{X}, w \sim \{0,1\}^m} & L_{acc}(w, \hat{w}; E, D) \\ & + \lambda (L_{con}(\mathbf{x_o}, \mathbf{x_w}; E) + L_{emb}(\mathbf{e_o}, \mathbf{e_w}; E)) \end{aligned} \tag{6}$$

$$L_{acc}(w,\hat{w};E,D)=\frac{1}{n}\sum_{k=1}^{n} w_k \log \hat{w}_k + (1-w_k)\log(1-\hat{w}_k) \quad (7)$$

$$L_{con}(\mathbf{x_o},\mathbf{x_w};E)=\left\|\mathbf{x_w}-\mathbf{x_o}\right\|_2^2 \quad (8)$$

$$L_{emb}(\mathbf{e_o},\mathbf{e_w};E)=\left\|\mathbf{e_w}-\mathbf{e_o}\right\|_2^2 \quad (9)$$

$$\mathbf{x_w}=E(\mathbf{x_o},w)+\mathbf{x_o} \quad (10)$$

$$\dot{\mathbf{x}}_\mathbf{w}=\mathbf{x_w}+noise \quad (11)$$

$$\hat{w}=D(\dot{\mathbf{x}}_\mathbf{w}) \quad (12)$$

where $\mathbf{e_w}$ is the embedding of the watermarked data $\mathbf{x_w}$ , $w_k$ and $\hat{w}_k$ are the $k^{th}$ bit of the input watermark and recovered watermark, respectively.

The loss function of the watermarking network includes 3 terms: the accuracy loss ($L_{acc}$), the content loss ($L_{con}$) and the embedding-matching loss ($L_{emb}$), as shown in (6) – (9). $\lambda$ is the balancing weight. $L_{acc}$ represents the binary cross-entropy, which guides the decoder to recover the watermark embedded by the encoder. $L_{con}$ employs mean squared error to minimize the point-to-point discrepancies between $\mathbf{x_o}$ and $\mathbf{x_w}$ to make the watermark invisible. $L_{emb}$ minimizes the divergence between $\mathbf{e_o}$ and $\mathbf{e_w}$ to further enhance the invisibility of the watermark, as $\mathbf{e_o}$ and $\mathbf{e_w}$ represent high-level features of $\mathbf{x_o}$ and $\mathbf{x_w}$ extracted by the neural network.

Since improving the watermark decoding accuracy may impair the watermark invisibility and increase the distortion to the original data, this paper adopts a dynamic weighting training strategy. The hyper-parameter $\lambda$ in (6) is initially set to 0 at the beginning of training to ensure that the model focuses on the accuracy of watermark reconstruction. Once the accuracy reaches a predefined threshold, $\lambda$ starts to increase with training iterations and eventually stabilizes. This allows the model to continue improving watermark invisibility while maintaining the watermark to be recognizable.

### 3.3 Noise Layer

In real-world scenarios, energy system time series data may suffer from various distortions such as missing data caused by communication failure, noise injection due to electromagnetic interferences, etc. Such distortions will alter the watermarked data and bring challenges to the watermark recognition. To enhance the robustness of watermark under data distortion scenarios, a noise layer is introduced before the decoder to simulate real-world disturbances. The noise layer injects Gaussian noises with 0 mean and 0.1 standard deviation to the watermarked data $\mathbf{x_w}$. The decoder is then responsible for recognizing the watermark from the noised data $\dot{\mathbf{x}}_\mathbf{w}$.

### 3.4 Watermark Encoding Based on ASCII

Watermarks typically need to carry clear and meaningful information to establish the copyright ownership rather than being a random bitstring. For this purpose, we chose ASCII, a widely used encoding system, to create binary bitstring encodings for the text-based watermarks. In the ASCII system, each character corresponds to a unique decimal value in the ASCII table. For example, letter A corresponds to 65. This decimal value is then converted into an 8-bit binary number, so A becomes 01000001. Following this rule, the text-based watermark can be converted to a binary bitstring that is ready to be embedded into the original data.

## 4. Case Study

To evaluate the effectiveness of the proposed watermarking method, we utilize the 1-minute resolution smart meter dataset from Pecan Street, Austin, TX to formulate the test case [36]. 704 residential daily load profiles are selected to formulate the time series data asset to be protected. A binary bitstring with length 100 is set up as the watermark to be embedded. As a comparison, a time-domain watermarking model is also tested with the same configuration as the proposed frequency-domain method but without the frequency-domain data preprocessing process. Additionally, two conventional watermarking techniques, specifically the Least Significant Bit (LSB) method and the Discrete Wavelet Transform (DWT)-based approach, are implemented as baselines.

### 4.1 Invisibility

Two daily load profiles before and after watermarking are shown in Fig. 6 as an example. We can see that the modification of watermarking process to the original data is almost visually indistinguishable. To further quantify the invisibility of the proposed watermarking method, the following indexes are calculated between the original and the watermarked data: Root Mean Squared Error (RMSE), Fréchet Inception Distance (FID), Cosine Similarity (CS), Spectral Similarity (SS) and Kullback–Leibler divergence (KL). For RMSE, FID and KL, smaller value indicates better invisibility. For CS and SS, closer to 1 means better invisibility. As a comparison, we add Gaussian noise with different standard deviations to the original dataset to simulate different levels of modifications. Results are shown in Table IV.

In Table IV, time- and frequency-domain watermarking models achieve satisfying invisibility, resulting in minor modifications to the original dataset. Particularly, the frequency-domain watermarking model achieves the best performances on all metrics, demonstrating the effectiveness of the proposed frequency-domain data preprocessing method. The distortion of the frequency-domain watermarking to the original dataset is equivalent to the noise injection with std = 0.001. Such tiny data distortion is considered negligible during the data sharing process for research purposes, such as data analytics, parameter identification and machine learning model training, etc.

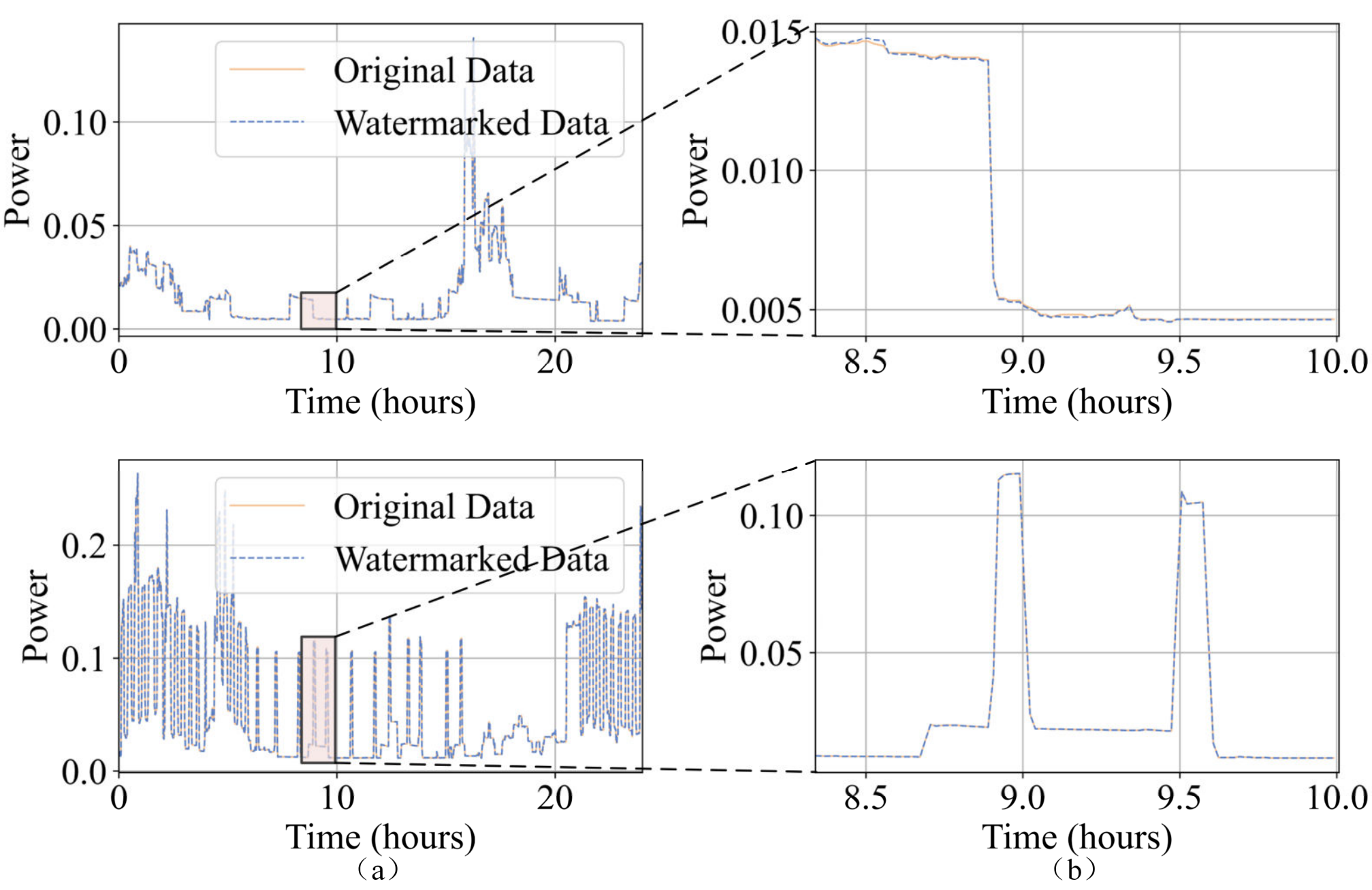

Fig. 6. Comparison of samples of the original data and the watermarked data. (a) Results overview on the samples, (b) regional zoom-in of the results.

Table IV Model metrics

| Models | Metrics | | | | |
|---|---|---|---|---|---|
| | **RMSE** | **FID** | **CS** | **SS** | **KL** |
| LSB | $1.113\times10^{-4}$ | $1.308\times10^{-4}$ | 0.9999 | 0.9999 | 0.0002 |
| DWT | $5.825\times10^{-2}$ | 0.1907 | 0.9944 | 0.9944 | 0.2563 |
| Time-domain | $5.889\times10^{-5}$ | $3.317\times10^{-7}$ | 0.9934 | 0.9967 | 0.0149 |
| **Frequency-domain** | $\mathbf{4.285\times10^{-9}}$ | $\mathbf{1.287\times10^{-11}}$ | **0.9999** | **0.9999** | $\mathbf{3.383\times10^{-6}}$ |
| Noise(std=0.1) | $1.001\times10^{-2}$ | $3.598\times10^{-3}$ | 0.5556 | 0.6923 | 5.7164 |
| Noise(std=0.05) | $2.505\times10^{-3}$ | $3.907\times10^{-4}$ | 0.8008 | 0.8722 | 4.9715 |
| Noise(std=0.01) | $1.002\times10^{-4}$ | $8.722\times10^{-7}$ | 0.9889 | 0.9945 | 3.1666 |
| Noise(std=0.005) | $2.504\times10^{-5}$ | $5.172\times10^{-8}$ | 0.9972 | 0.9986 | 1.5731 |
| Noise(std=0.001) | $9.998\times10^{-7}$ | $6.004\times10^{-13}$ | 0.9998 | 0.9999 | 0.0085 |

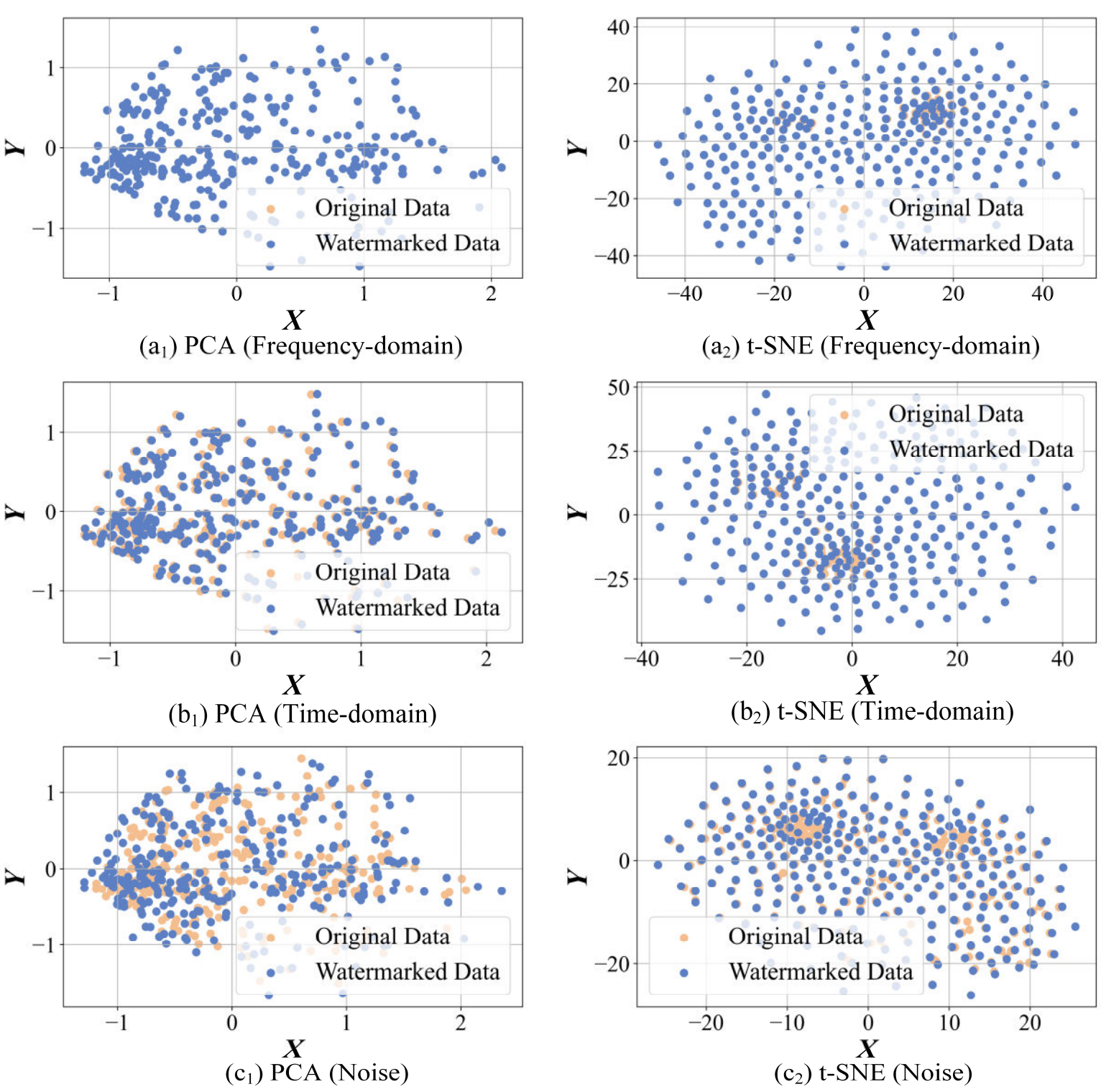

Fig. 7. 2-D visualizations of 300 samples before and after watermarking, based on PCA and t-SNE. (a) Frequency-domain watermarking, (b) time-domain watermarking, (c) noise injection with std = 0.1.

We select 300 samples among the total 704 daily profiles and plot out their 2-dimensional distribution to give an overview regarding invisibility of the proposed watermarking method, shown as Fig. 7. Principle Component Analysis (PCA) and t-Distributed Stochastic Neighbor Embedding (t-SNE) are implemented for dimension reduction purposes. Accordingly, the distribution of frequency-domain watermarked samples is highly overlapped with the original samples, again indicating the superior invisibility of the proposed method. The time-domain watermarking results show a nonnegligible systematic bias cause by the frequency bias issue mentioned in Section II.B, making the watermark risky to be detected. Both the frequency- and time-domain models show better invisibility and less data distortion than noise injection with std = 0.1.

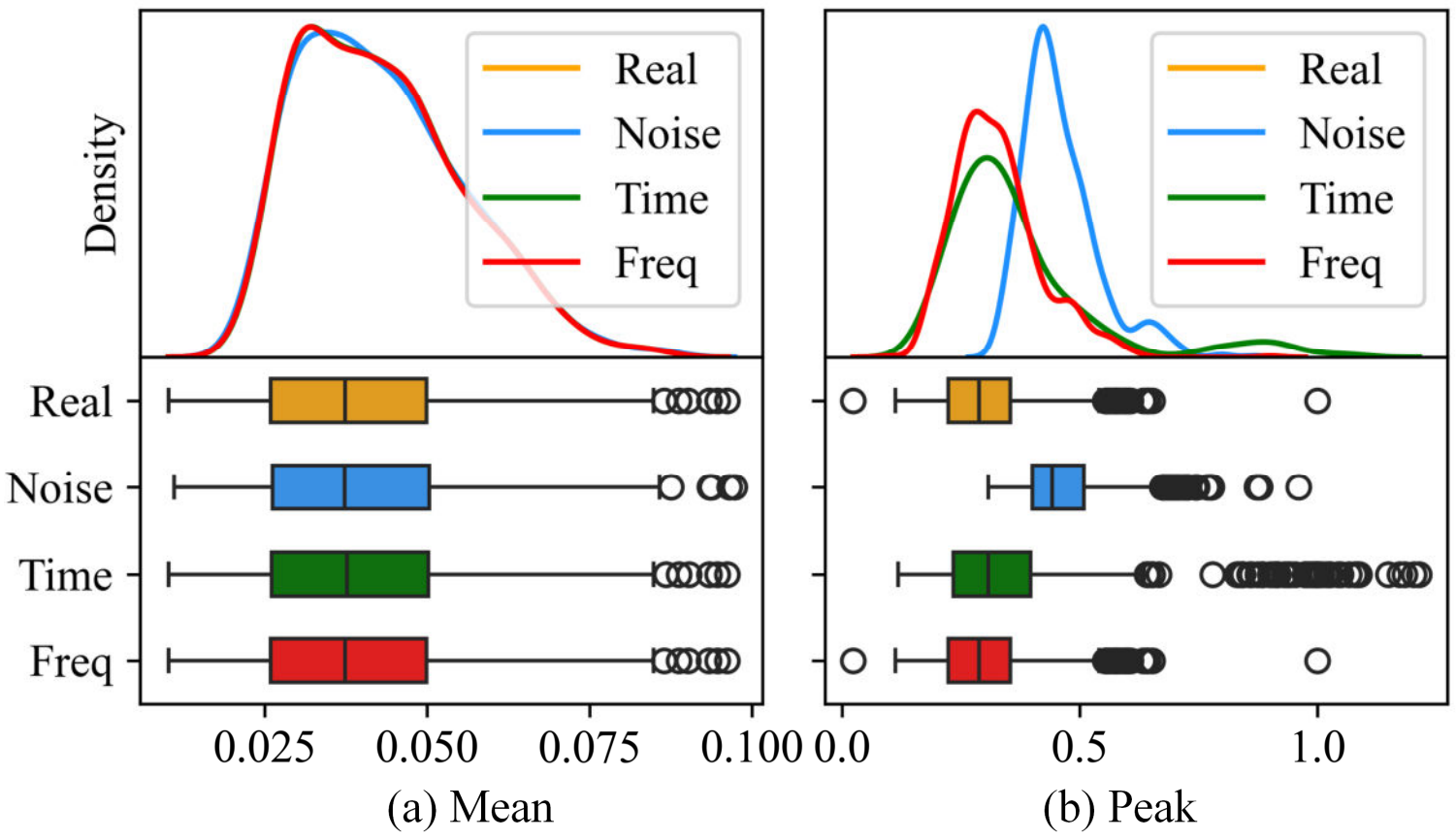

Fig. 8. (a) Mean probability density distributions and boxplots, (b) peak probability distributions and boxplots.

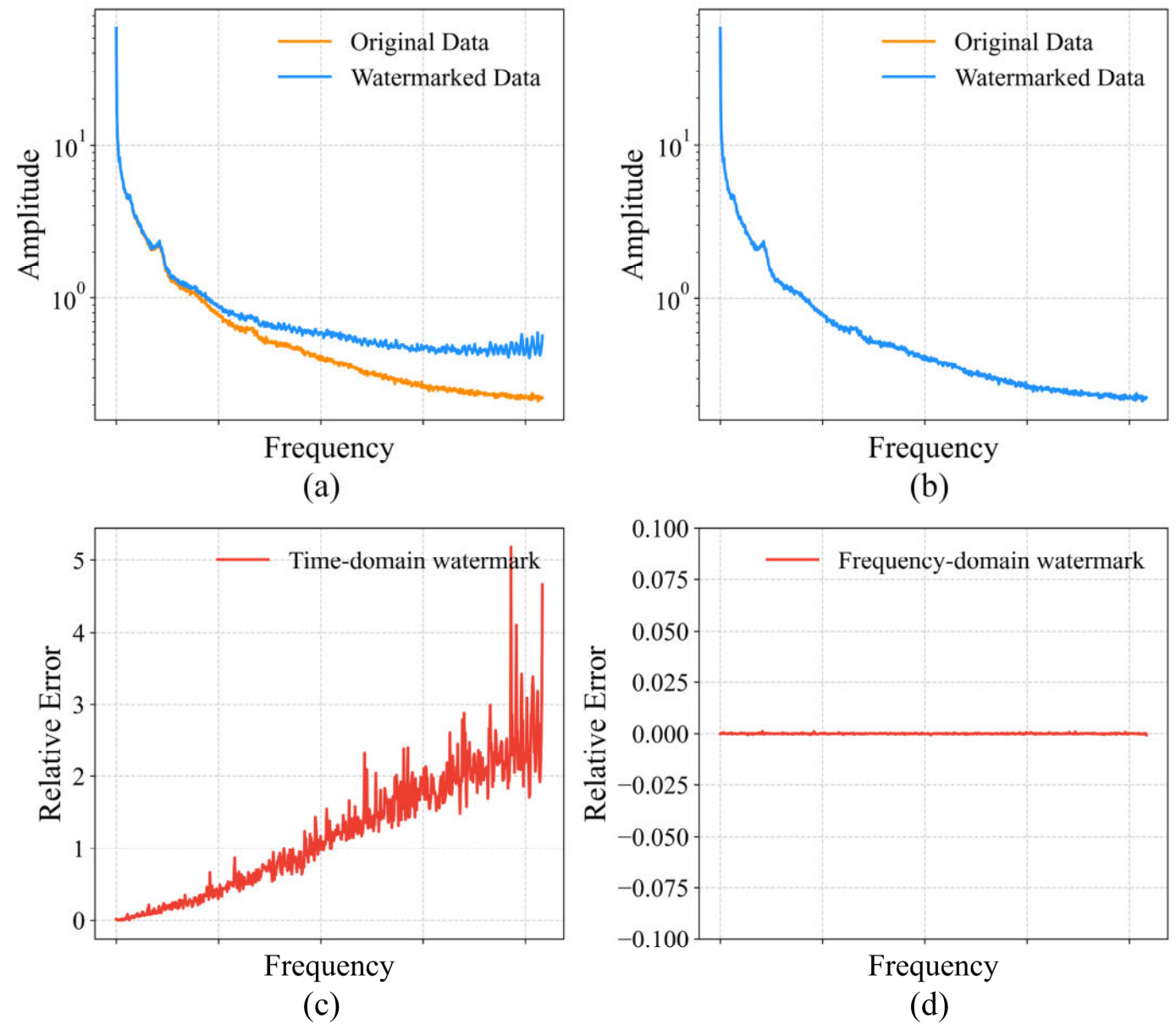

Fig. 9. Averaged fitting errors on different frequencies. (a) and (c) show the time-domain model performance, while (b) and (d) show the proposed frequency-domain model performance.

Considering a statistical perspective, we present the probability density distributions and boxplots for the mean and peak values of real data, noised data, time-domain watermarked data and frequency-domain watermarked data, as shown in Fig. 8. The mean distributions across all data types are closely aligned. However, in terms of peak value distributions, the noised data displays a notable upward bias, while the time-domain watermarked data exhibits long-tail bias. The frequency-domain watermarked data remains highly similar to the real data, again demonstrating superior invisibility.

To demonstrate the effectiveness of the proposed frequency-domain method in correcting the frequency bias issue, Fig. 9 compares the errors of time- and frequency-domain models in different frequencies. Here, the errors of time-domain models increase significantly along with the frequency, showing evident frequency bias which impairs the watermarking invisibility. On the contrary, the proposed frequency-domain method has almost zero errors across all frequencies, successfully eliminating the frequency-bias issue.

## 4.2 Restorability and Robustness

We evaluate the restorability and robustness of the proposed watermarking method by calculating the bitwise accuracy between the original and recovered watermarks under two perturbation scenarios: 1) Noise injection with different standard deviations and sample proportions. 2) Missing data by masking different ratios of samples perturbations are implemented to the watermarked samples to see whether the decoder can still recognize and restore the embedded watermarks. Note that 50% accuracy means the watermark cannot be restored, indicating the decoder is making random guessing for the binary watermark. Experiment results are shown in Fig. 10. Using Fig. 10, we reach the following observations:

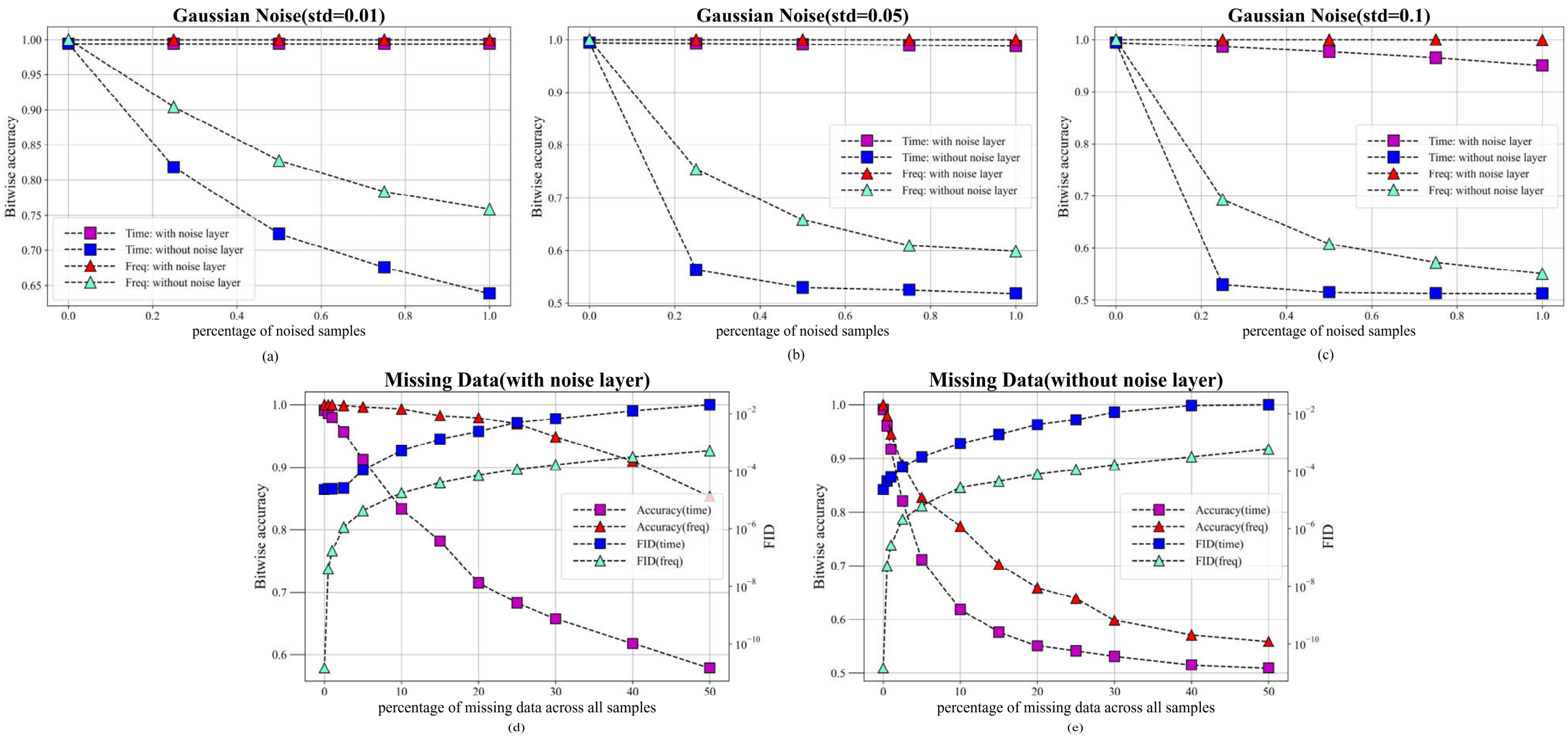

Fig. 10. Reconstruction performance of the decoder under different noise injection levels and missing data ratios. (a) Noise std = 0.01, (b) noise std = 0.05, (c) noise std = 0.1, (d) with noise layer under different missing data ratio, (5) without noise layer under different missing data ratio.

1) Both time- and frequency-domain models can achieve 100% watermark restoration accuracy under no-perturbation scenarios, demonstrating that the proposed watermarking framework has good restorability.

2) The proposed frequency-domain model can function well under severe data conditions. For noise injection scenarios in Figs. 10(a)-(c), the frequency-domain model remains above 99% watermark restoration accuracy, showing superior noise resistance. For missing data scenarios in Figs. 10(d)-(e), the frequency-domain model achieves 98.26% watermark restoration accuracy when the missing data ratio reaches 15%, even though the data quality had significantly degraded. At 50% missing data rate, the frequency-domain model can still achieve 85% accuracy, given the severe data quality degradation. Such observations demonstrate that the watermarking model, especially the proposed frequency-domain model with noise layer, is robust and can function well under severe data conditions.

3) The introduced noise layer significantly improves the watermarking robustness. As shown in Figs. 10(a)-(c), the accuracy of the watermarking model without noise layer drops significantly with increasing the percentage of noised samples. On the contrary, the frequency- and time-domain models with noise layers remain above 95% accuracy even when all training samples are polluted with strong Gaussian noises with standard deviation 0.1. Similar observation can be seen in Figs. 10(d)-(e). The accuracy of the watermarking models with noise layers decreases with the increasing missing data ratio. Such observations demonstrate the value of noise layer in enhancing the model robustness to resist the influence of data perturbation.

To evaluate the robustness, we define successful watermark detection as achieving a verification bit rate of at least 75% for each sample (with *p-value* = $2.8\times10^{-7}$). The successful rates under various perturbation scenarios are summarized in Table V. The LSB and DWT methods embed watermarks into individual time-domain or frequency-domain bits, leading to rapid accuracy degradation when data changes. In contrast, our proposed method demonstrates superior robustness. Table V shows that while LSB and DWT are highly sensitive to noise and missing data, our approach maintains high successful detection rates even under challenging conditions.

Table V Successful detection rates (%)

| **Models** | **Scenarios** | | | | |
|---|---|---|---|---|---|
| | Noise(std=0.001) | Noise(std=0.01) | Noise(std=0.05) | Noise(std=0.1) | Missing(50%) |
| LSB | 0 | 0 | 0 | 0 | 59.63 |
| DWT | 100 | 98.6 | 0 | 0 | 9.21 |
| Time-domain | 100 | 100 | 100 | 100 | 74.01 |
| **Frequency-domain** | **100** | **100** | **100** | **100** | **84.09** |

### 4.3 Secrecy

Watermarking should not be easily detected by third parties as it could lead to watermark manipulation or removal. In this section, we assume three levels of attackers who attempt to identify whether there exists a watermark embedded in the dataset (with or without watermark), according to different information availability.

1) Weak attacker is assumed to have access to the structure of watermarking model, but cannot access to the trained model parameters or the true watermark.

2) Moderate attacker is assumed to have access to both the watermarking model structure and the true watermark, but cannot access the trained model parameters.

3) Strong attacker is assumed to have full access to all the watermarking model structure, parameters, and the true watermark.

The detection process is typically framed as a binary classification problem to distinguish between watermarked and unwatermarked samples, named as the ATS (Artificial Training Sets) method [37]. However, because in practice there are no ground-truth labels for the watermark existence from attackers' perspective, the detection process is conducted in an unsupervised manner. The attackers will implement the watermarking algorithm to the original dataset they obtained to generate set **A** containing the original data and the single-watermarked data. Then **A** will be fed into the watermarking model again to obtain **B** containing single- and double-watermarked data, **C** containing double- and triple-watermarked data. **C** is used as the positive dataset while **A** serves as the negative dataset to train an SVM classifier, and **B** will be used as the testing set to evaluate the trained SVM classifier. Once the classifier can make effective classification on B, the attacker can distinguish between the single- and double-watermarked samples, thus recognize the existence of watermark to make further threats.

Note that attackers might need to follow different procedures to conduct the above ATS detection process. Strong attackers can directly conduct ATS because they have access to the trained model and the true watermark. Moderate and weak attackers need to train the watermarking model by themselves, while weak attackers also need to generate a synthetic watermark as the model input, leading to stochastic performances. To alleviate the randomness, five models with different initializations will be trained to create ATS, while the sixth model will be created to test the classification.

Simulation results are still based on the 704 residential load profile datasets embedded with time- and frequency-domain watermarks, which are used as the dataset disclosed to the attackers. The classification accuracy of attackers is summarized in Table VI. We can see that only strong attackers have chances to detect the watermark (accuracy > 50%), while both weak and moderate attackers completely fail (accuracy $\approx$ 50%). Especially on the frequency-domain watermarked dataset, even the strong attackers cannot make effective detection. Note that in practice, attackers usually cannot access any information about the watermarking model or the true watermark, making the detection even more challenging. Such results demonstrate that the proposed frequency-domain watermarking method is sufficiently secure and secret to prevent from being detected by attackers.

Table VI Classification accuracy under attacks

| Model | Attacker | Access to model | Access to watermark | Access to parameter | Classification Accuracy |
|---|---|---|---|---|---|
| Time-domain | Weak | Yes | No | No | 0.4978 |
| | Moderate | Yes | Yes | No | 0.5099 |
| | Strong | Yes | Yes | Yes | **0.9212** |
| Frequency-domain | Weak | Yes | No | No | 0.5014 |
| | Moderate | Yes | Yes | No | 0.5035 |
| | Strong | Yes | Yes | Yes | **0.5545** |

## 4.4 False-positive Detection

False-positive detection is defined as the model detecting the existence of watermarks on data samples who do not contain one. False-positive detection may lead to misjudgments on data ownership and cause potential copyright disputes. In this paper, we define a positive detection when a decoder can successfully restore over 75% bits in a watermark [18]. Accordingly, the positive detection rate is defined as the ratio of the number of the detected watermarked samples divided by the total number of watermarked samples. A well-designed watermarking model must ensure a low false-positive rate. In this section, we conduct false-positive detection experiments by feeding the original 704 samples and 704 watermarked samples to the trained decoder to calculate the positive detection rate, shown in Table VII. Accordingly, the watermarking model achieves perfect detection with zero detection errors.

Note that the watermark $w$ used in this paper is a bitstring with 100 digits, leading to $2^{100}$ possible combinations. Such a large solution space can effectively avoid the chances for the watermark of being coincidently detected, which may lead to bias on the positive detection rate.

Table VII Positive detection rate

| Dataset | Original | Watermarked |
|---|---|---|
| **Positive detection rate** | 0% | 100% |

## 4.5 Generalization

The proposed watermarking method should be able to be generalized to other energy system time series datasets beyond the load profiles studied in this paper. To this point, we build a photovoltaic power generation dataset including 224 daily power output profiles of a solar station in Australia to test the effectiveness of the proposed method.

We employ a pre-training and fine-tuning strategy: the model, which has already been pre-trained on load data, requires only fine-tuning to adapt to the photovoltaic (PV) dataset. Results are shown in Fig. 11 which demonstrate the invisibility, and fine-tuning model can also achieve 100% watermark restoration accuracy. We can see that the model also performs well in the photovoltaic power output dataset.

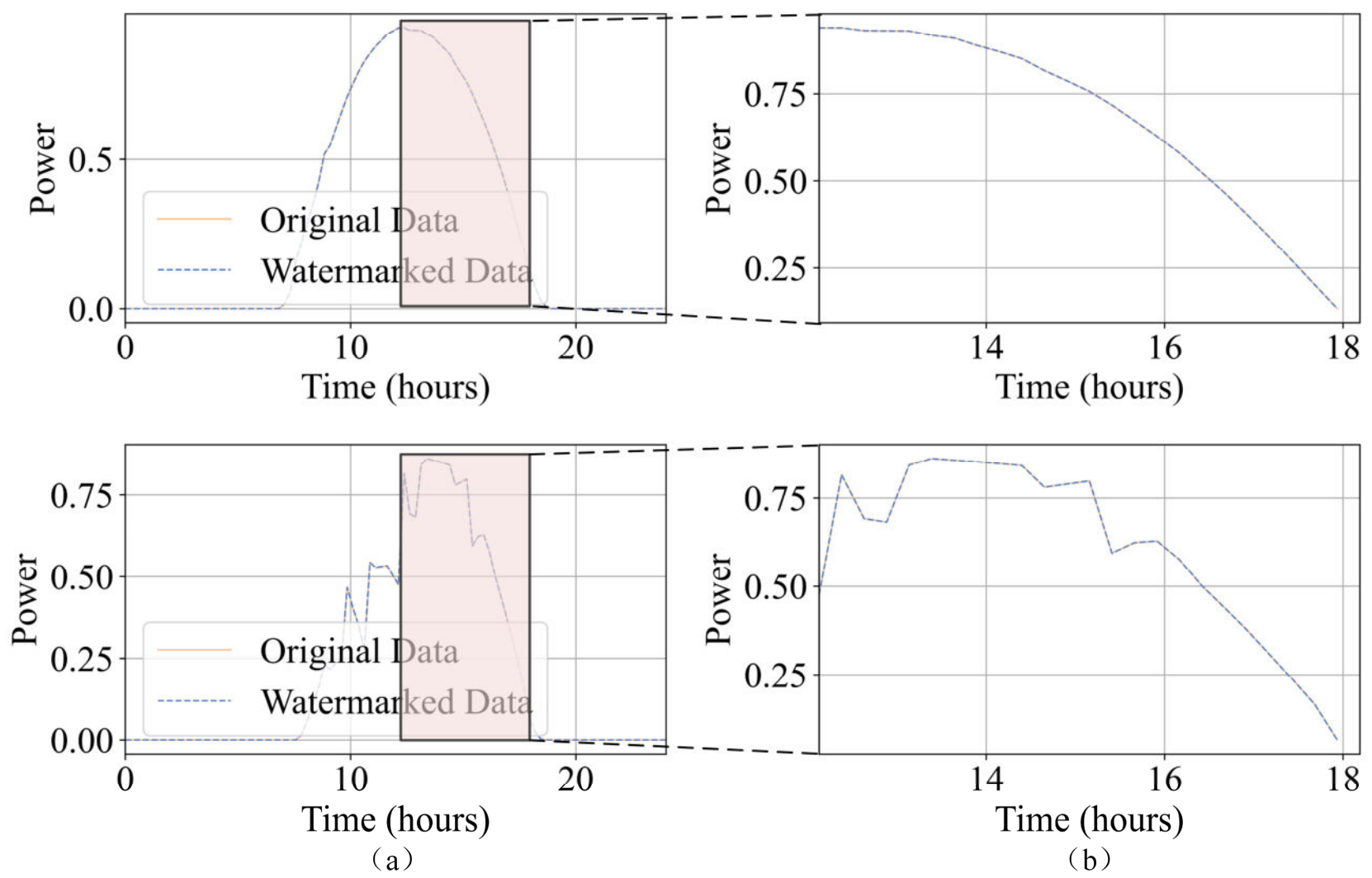

Fig. 11. Comparison of samples of original data and watermarked data. (a) Results overview on the samples, (b) regional zoom-in of the results. Data source: DKASC[38].

### 4.6 Capacity

For the data asset protection purposes, the watermark should have sufficient capacity to embed enough information into the dataset. However, larger information embedding will cause more data distortion to the original dataset and increase the difficulty of restoring the watermark, thus impairing the watermark invisibility and restorability. We test 5 watermarking models with watermark length 100, 500, 1000, 2000, 3000, and plot out the watermark detection accuracy and FID scores reflecting data distortion, shown in Fig. 12.

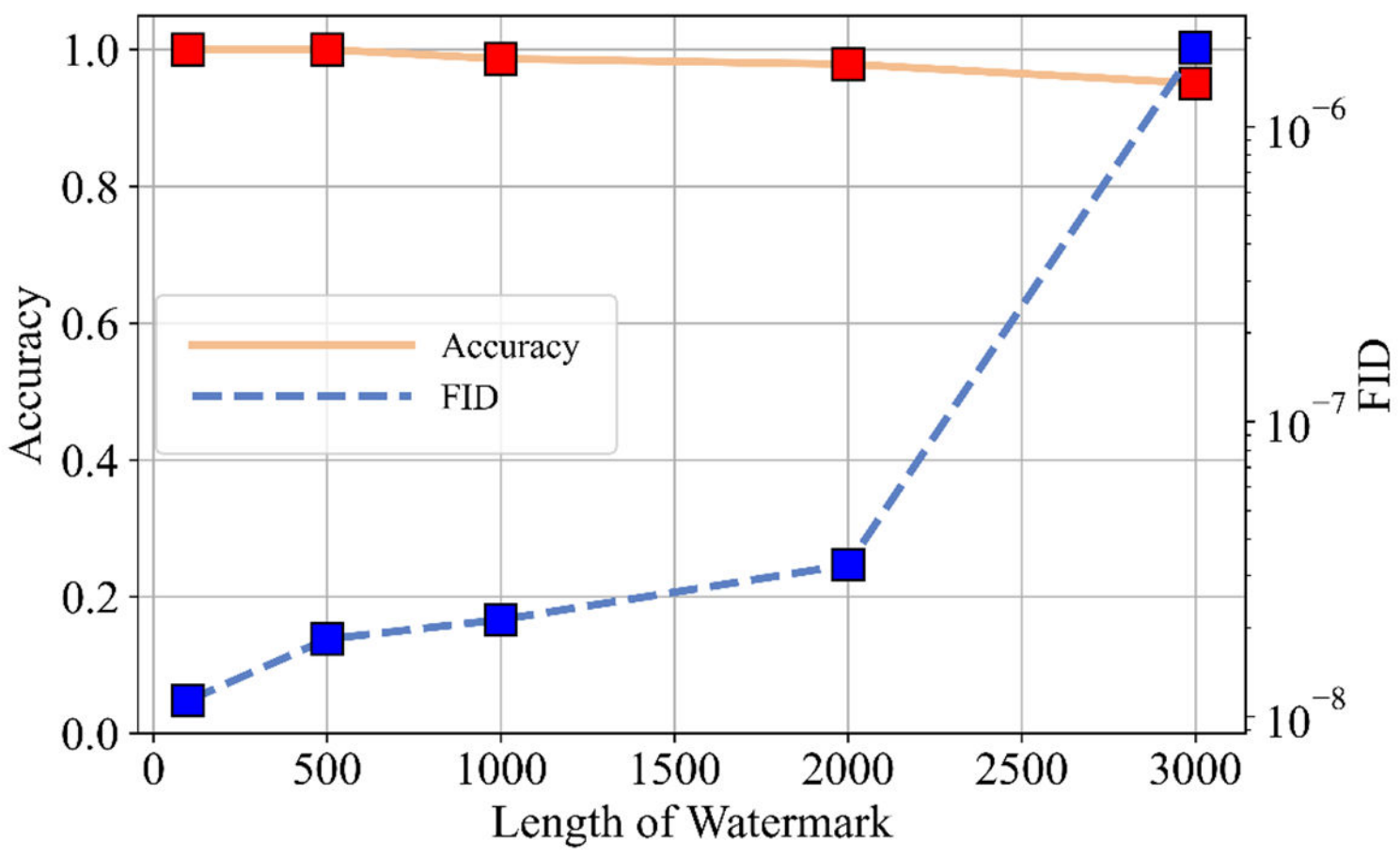

Fig. 12. Watermark detection accuracy and FID scores with different watermark lengths.

We can see that the watermark detection accuracy decreases while the FID score increases, with the increasing of watermark length, indicating the deteriorated invisibility and restorability. However, we also notice that when the watermark length reaches 3000 (equal to 375 characters), the accuracy is still above 95% and the FID remains at a low level, meaning that the proposed watermarking method is tolerant for large-scale information embedding. When implementing watermarking in practice, the data owners are suggested to select a smaller watermark length as long as the watermark has enough capacity for embedding the intended information.

## 5. Discussion

The watermarking method introduced in this paper has been rigorously evaluated across several critical dimensions: invisibility, restorability, robustness, secrecy, false-positive detection rate, capacity, and generalization. High-level invisibility is fundamental to the application of watermarking technologies, entailing only negligible alterations to the data, as illustrated in Fig. 6, 7, 8, and 9, and Table IV. Superior restorability and robustness are paramount for watermarking to serve as a reliable method of ownership verification, ensuring that watermarks remain detectable even after intentional modifications by adversaries, as depicted in Fig 10 and Table V. Secrecy ensures that watermarks are difficult for attackers to locate or remove, as shown in Table VI. A low false-positive detection rate guarantees the reliability of watermark assertions, as presented in Table VII. Adequate capacity allows the watermark to securely embed sufficiently long keys, as detailed in Fig. 11, while good generalization ensures the watermark can be readily adapted to various datasets, as demonstrated in Fig. 12. Consequently, this watermarking approach presents a viable solution for verifying dataset ownership. However, the necessity of acquiring the suspicious dataset as input for the decoding process somewhat restricts the applicability of this method. In fact, if an attacker merely misappropriates the data without publicly exploiting it for additional illegal gains, it becomes challenging for the data owner to ascertain the unauthorized usage definitively. In summary, despite the limitation, the proposed methodology offers a credible means of energy system time series data asset ownership verification.

## 6. Conclusion

In this paper, a deep-learning-based watermarking method is introduced to protect the energy system time series dataset. The proposed model employs an encoder-decoder structure with a comprehensive loss function composed of accuracy loss, content loss and embedding-matching loss to achieve optimal balance between watermark invisibility and restorability. A frequency-domain data preprocessing method is proposed to eliminate the frequency bias issue during neural network training, and a noise layer is introduced to the watermarking network to enhance the model robustness. Case study based on real-world load and photovoltaic dataset demonstrates that the watermarking model has negligible influence on the original dataset and is considered invisible. Meanwhile, the watermark is robust to severe data distortion, can only be restored by the owner but remains undetectable to the unauthorized third parties. Therefore, the proposed watermarking method is considered effective and could be a promising way of protecting energy system data assets.

Future work may focus on extending the application of the watermarking method to protect other types of energy system data, such as system topology, equipment models, etc.

## 7. Acknowledgement

This work was supported by National Natural Science Foundation of China under Grant 52307121, and also supported by Shanghai Sailing Program under Grant 23YF1419000.